\documentclass[a4paper,11pt]{article}

\usepackage{jheppub} 

\usepackage[T1]{fontenc} 

\title{Non-perturbative renormalization: a new way}

\author{G.~B.~ Pivovarov}
\affiliation{Institute for Nuclear Research of the Russian Academy of Sciences,\\Prospect 60-letiya Oktyabrya, 7a, 117312, Moscow, Russia}

\emailAdd{gbpivo@ms2.inr.ac.ru}

\abstract{The notion of non-perturbative renormalization is discussed and extended. Within the extended picture, a new non-perturbative representation for the generating functional of Green functions of quantum field theories is suggested. It is argued that the new expression agrees with the standard renormalized perturbation theory if the latter is renormalized in an appropriate renormalization scheme.}

\begin{document} 
\maketitle
\flushbottom

\section{Introduction}
\label{Introduction}

Within the version of non-perturbative renormalization based on functional integral \cite{Glimm:1987ng},
generating functional of the connected Green functions $W(J)$ is related to the bare action functional $S_\Lambda(\phi)$ with the functional Fourier transform:
\begin{equation}
\label{Fourier}
e^{W(J)}=\mathcal{F}_\Lambda[S_\Lambda](J)\equiv\int\mathcal{D}_\Lambda\phi\,\,e^{-S_\Lambda(\phi)-iJ\phi}.
\end{equation}
Here $\Lambda$ is an UV-cutoff. The action depends on the cutoff, as well as the functional measure $\mathcal{D}_\Lambda\phi$. The task of the non-perturbative renormalization is to find such a dependence of $S_\Lambda(\phi)$ on $\Lambda$ that the left hand side of eq. (\ref{Fourier}) would be finite in the continuum limit $\Lambda\rightarrow\infty$, and, on top of that, the resulting Green functions would comply with a certain set of axioms, like locality and Lorentz covariance. 

In a more broad context, non-perturbative renormalization is a term used in constructive quantum field theory (see \cite{Mastropietro:2008zz}). The latter is motivated with the need to put the phenomenological successes of quantum field theory on a firm mathematical ground (for an overview of these attempts see \cite{Summers:2012vr}). This is not my motivation. I am willing to accept that the Nature is described with the asymptotic series of the perturbation theory.

Not a lack of rigor in the perturbation theory is the major problem of the present day high energy physics, but a lack of model selectivity: any renormalizable theory would do from the standpoint of the perturbation theory. The plethora of field components in the Standard model remains to be a complete mystery.

My motivation in looking for a non-perturbative renormalization is to find the missing selectivity, that is, I hope that some particular list of fields may turn out to be more amenable to a non-perturbative renormalization than the other, despite the fact that they are both completely acceptable from the standpoint of the perturbation theory.

Compliance of Green functions with a list of axioms is not my subject. I will check my construction comparing its outcome with the standard perturbation theory.  
I will suggest an object---I call it quantum transform---which may have a meaning beyond perturbation theory, and whose expansion agrees with the renormalized perturbation theory for Green functions.

I suggest to extend in a pragmatic direction the program of the functional integral non-perturbative renormalization. Indeed, the program related to eq. (\ref{Fourier}) insists on the particular roles played in eq. (\ref{Fourier}) by the bare action and the functional integral. Pragmatically, even if one would succeed in finding the ``right'' dependence of $S_\Lambda(\phi)$ on $\Lambda$, one would not be done with the ``renormalization'', because the result in the left-hand side would be parametrised with a finite number of renormalized couplings involved in the bare action $S_\Lambda(\phi)$. So, to complete the renormalization, one would have to express the renormalized couplings in terms of some physically observable parameters. 

This may seem to be a technicality, because trading the renormalized couplings for observable ones is only a change of parametrisation involving only a finite number of parameters. On the other hand, it may be considered as a matter of principle for a theory involving parameters that cannot be taken from a more fundamental theory. And the description of an object in terms of parameters extracted from itself is a natural source of the renormalization group, which results from the requirement that the outcome of the description would be independent of the way the parameters used in the description have been extracted from the object under description.
 
I take in this paper namely the second point of view, and will describe the generating functional of connected Green functions $W(J)$ not in terms of the action $S_\Lambda(\phi)$, but in terms of a functional $I_P(\phi)$, which I will mockingly call the inaction functional. The inaction functional is extracted from $W(J)$ in an explicitly given way, and is independent of the cutoff. It contains controllable number of parameters, not less than the action $S_\Lambda(\phi)$, but they are not some theoretical renormalized couplings. Instead, they are parameters extracted in a predefined way from $W(J)$. There is an arbitrariness in this extraction parametrised with some $P$ to be defined (see below). 
Comparing again action to inaction, while the inaction keeps like the action the parameters of the theory, instead of unknown dependence on the cutoff $\Lambda$, it has a predefined dependence on some $P$ (still to be defined).

The rest of the information about the Green functions that is not captured by the inaction functional is sent to another functional of fields, $V_P(\phi)$. Intuitively, $V_P(\phi)$ contains the connected vertexes with external legs amputated and local parts of non-negative mass dimensions removed. I call it the vertex functional. 

The task of the theory is to express the vertex functional in terms of the inaction functional:
\begin{equation}
\label{my-transform}
V_P(\phi)=\mathcal{Q}_P[I_P](\phi)
\end{equation}
Here $\mathcal{Q}_P$ stands for \emph{quantum transform}. It replaces the functional Fourier transform of eq. (\ref{Fourier}). Comparing this formula to eq. (\ref{Fourier}) I point out that dependence on the cutoff $\Lambda$ is replaced here with the dependence on the extraction
procedure, $P$, the action is replaced with the inaction, and the generating functional of the connected Green functions  with the vertex functional. I call this formula the inaction representation. Within the inaction formulation, UV divergences are considered as an artefact of the formulation related to the representation of eq. (\ref{Fourier}). An UV regularization can be used at intermediate stages in constructing the quantum transform, but it should be well defined in the continuum limit, because both the inaction and the vertex functionals it relates to one another are finite by construction, and do not depend on the UV cutoff.

It is possible to combine the vertex functional and the inaction functional back into the generating functional of the Green functions $W(J)$. Requiring its independence on the extraction procedure gives some renormalization group equations. Perturbatively they were considered in \cite{Pivovarov:2009wa} for $\phi^4$. It was concluded that, under certain definition, the running mass of a scalar field has a Landau pole.
The inaction approach was further developed in 
\cite{Pivovarov:2012wz}
to include gauge symmetries. In the present paper, I sharpen the formulation paying special attention to independence of the perturbation theory.

Below I will do my current best to give a definition to the quantum transform $\mathcal{Q}_P$ in eq. (\ref{my-transform}), and will check that it makes sense in all orders of perturbation theory. This definition is ``derived'' from the basic representation eq. (\ref{Fourier}). But since the functional Fourier transform is not easy to define, the following definition of the quantum transform may as well be postulated. The definition of the quantum transform I give in Section \ref{The quantum transform} employs functional integration. So, it requires a regularization. But, in contrast to the functional Fourier transform of eq.(\ref{Fourier}), the quantum transform should exist in the continuum limit. So, the regularization and the functional integral are only means used presently to define it. A future formulation should postulate some defining properties of the quantum transform, and build the theory around them. These defining properties are still to be discovered.

One of the merits of the inaction approach is that one can change the extraction procedure at will, obtaining various kinds of separations between the information one takes as known, and the information to be yielded by the theory. An example of this sort will be given below: suppose I know the $N$- point Green functions up to some $N$, and want to express the rest of Green functions in terms of the known ones. There are quite interesting Feynman rules emerging within the inaction approach to solve this task. What is interesting about these rules is the mechanism the UV divergences take care about themselves. From this perspective, UV problem looks like a problem of computing some combinatorial factors: the combinatorics gives zero factors by the UV divergent diagrams. This point is pushed further in \cite{Kim&Pivovarov}.   

Another interesting subject appearing in the context of this paper is a comparison of the inaction representation eq. (\ref{my-transform}) with the known exact results, in particular, with the results on the zero-dimensional field theory (see, for example, \cite{Argyres:2001sa}). A work in this direction is under way.

The rest of the paper is organized as follows. In the next section, I detail the setup, explaining all the big letters I used above to introduce my objects. Then, in Section \ref{The quantum transform} I define the quantum transform and see how it works in perturbation theory. In Section \ref{Combinatorics} I consider the funny way the UV diveregences disappear if one uses the inaction approach to express, say, a six-point Green function in terms of four- and two- point Green functions. In the last section, I list again the merits and demerits of inaction and questions it raises.

\section{The setup}
\label{Setup}

The $P$ labeling the extraction procedure in eq. (\ref{my-transform}) is a linear projector acting on field functionals, $P^2=P$. It projects out the part of functionals we consider to be known. Its relation to the conventional formulation is that it nullifies the bare action:
\begin{equation}
\label{projector}
P[S_\Lambda](\phi)=0.
\end{equation}
I will always assume that $P$ nullifies any constant functional:
\begin{equation}
\label{constant}
P(constant)=0.
\end{equation}

The requirements (\ref{projector}) and (\ref{constant}) generalize the properties of subtraction operator used in the BPHZ renormalization scheme.   In the context of perturbative renormalization, the subtraction operator acts on Feynman amplitudes (see, e.g., \cite{Collins:1984xc}). If I consider them as coefficient functions of field functionals, I obtain a projector acting on field functionals with the above properties. I will not rush with specifying explicitly the projector $P$. Instead, I list its properties required for my purposes. In this way I generalize the particular perturbative renormalizations considered previously. An example of a useful projector $P$ (a.k.a. subtraction operator), which, to my knowledge, have not been considered previously, see below in Section \ref{Combinatorics}.

The dimension of the space that $P$ nullifies (the dimension of its kernel), is the number of parameters one chooses to parametrise the theory with. In gauge theories, there are nonlinear relations between the couplings. The extraction procedure I discuss here does not capture these subtleties. I will parametrise the theory with the inaction functional belonging to the kernel of $P$. At  this stage $P$ only fixes the number of possible interactions (like three-gluon vertex and four-gluon vertex), but ignores that it may be necessary to restrict them to some nonlinear surface immersed in the kernel of $P$. (See more on this in \cite{Pivovarov:2012wz}.)

Next, I define the inaction functional and the vertex functional involved in eq. (\ref{my-transform}). For doing this, I want to amputate the external legs of the connected Green functions. For that I need a propagator. I take any invertible well-defined linear symmetric operator $D^{-1}_0$ acting on the fields and restricted by the condition
\begin{equation}
\label{propagator}
P[\phi D^{-1}_0 \phi]=0.
\end{equation}
So, the quadratic form of $D^{-1}_0$ is in the kernel of $P$, which is consistent with the above interpretation that $P$ projects out the known part of the theory, and I pretend that I know a good candidate propagator $D_0$. (I should warn the reader that I write all the signs for boson fields. The changes necessary to take into account the anti-commuting fields are straightforward. Another warning: my $W(J)$ does not have a linear term in the sources, so $J$ are the sources for the deviations of the fields from their vacuum values.) 

For me, the quadratic part of $W(J)$ deviating from $(-J D_0 J/2)$ will be due to interaction, and ``free'' theory has $W(J)$ exactly equal to $(-J D_0 J/2)$. Regardless of this interpretation, formally $P$ is an arbitrary projector, and $D_0$ is an arbitrary invertible propagator matrix satisfying condition (\ref{propagator}). And the condition (\ref{projector}) is not to be considered as restricting the projector $P$. It is the other way around: I will consider theories whose bare action satisfies the condition (\ref{projector}).

I drop the ``free'' part of $W(J)$ and amputate its external legs with the substitution $J\rightarrow iD_0^{-1}\phi$. This defines an ``amputee'' functional
\begin{equation}
\label{amputee}
A(\phi)\equiv W(iD^{-1}_0\phi)-\frac{1}{2}\phi D_0^{-1}\phi.
\end{equation}
The information in $A(\phi)$ is the same as in $W(J)$. It is the generating functional for connected vertexes with free part dropped and external legs amputated.

Now I define my two main characters: the vertex functional $V_P(\phi)$, and the inaction functional $I_P(\phi)$. Their sum is the above amputee $A(\phi)$. The inaction $I_P(\phi)$ is considered given, and $V_P(\phi)$ is to be determined in terms of $I_P(\phi)$. Accordingly,
\begin{equation}
\label{vertex}
V_P(\phi)\equiv P[A](\phi),
\end{equation}
and
\begin{equation}
\label{inaction}
I_P(\phi)\equiv (1-P)[A](\phi).
\end{equation}
So, the inaction is the part of the amputee in the kernel of $P$, and the vertex functional is the rest of the amputee.

A side note: it is always possible to tune $D_0$ in such a way that the quadratic part of the inaction disappears. It may be the most natural choice, but formally it is not required.

To define the quantum transform in the next section, I need two more ingredients. It is the $T$-product and the inverse $T$-product. They both are linear operators acting on functionals of fields. These operators are standard objects for generating the Feynman rules with the functional methods (see \cite{Vasiliev}). The definitions are as follows:
\begin{equation}
\label{t-prod}
T[F](\phi)\equiv e^{\frac{1}{2}\phi D^{-1}_0\phi} \sqrt{\det\frac{D_0}{2\pi}} \int\mathcal{D}_\Lambda J\,\,e^{-\frac{1}{2}JD_0J+iJ\phi} 
F(-iD_0J),
\end{equation}
and
\begin{equation}
\label{inverse}
T^{-1}[F](\phi)\equiv \frac{1}{\sqrt{\det(2\pi D_0)}}\int\mathcal{D}_\Lambda\phi^\prime\,\,\,e^{-\frac{1}{2}(\phi-\phi ^\prime)D^{-1}_0(\phi-\phi ^\prime)}
F(\phi^\prime).
\end{equation}
One checks that these operations are indeed inverse to one another. Also, there are the following representations for them:
\begin{eqnarray}
\label{representations}
T[F](\phi) &=& \exp{\big(-\frac{1}{2}\frac{\delta}{\delta\phi}D_0\frac{\delta}{\delta\phi}\big)}F(\phi),\\
T^{-1}[F](\phi)&=&\exp{\big(\frac{1}{2}\frac{\delta}{\delta\phi}D_0\frac{\delta}{\delta\phi}\big)}F(\phi).
\end{eqnarray}

Notice the asymmetry between eqs. (\ref{t-prod}) and (\ref{inverse}). This is due to implicit assumption that $D_0$ is positive definite. Formally, eq. (\ref{t-prod}) is obtainable from eq. (\ref{inverse}) with the substitution $D_0\rightarrow -D_0$, and a change of the integration variables which turns the integration into a Gaussian one (that is, with damping quadratic exponent), and removes the minus sign under the square root of the determinant.

I also stress that these $T$-products depend on UV regularization, and become singular operations in the continuum limit. To avoid clutter in the formulas, I don't make this evident in the notations, but keep in mind that $T$-products depend on the UV cutoff $\Lambda$.

The last requirement I need to impose on $P$ and $T$ is that $T$-products do not take a functional away from the kernel of $P$:
\begin{equation}
\label{requirement}
PT^{-1}(1-P) = 0.
\end{equation}
This is a natural requirement, because $(1-P)$ makes a functional local, $T^{-1}$ makes out of it a sum of tadpoles, which still keeps it local, thus $P$ nullifies it.

Anyway, to define the quantum transform in the next section, I need $P$, $T$, $T^{-1}$, and assume eqs. (\ref{projector}), (\ref{constant}) (\ref{propagator}), and (\ref{requirement}). With this at hand I can start constructing the quantum transform.

\section{The quantum transform}
\label{The quantum transform}

Take the $T$-product (\ref{t-prod}) of the exponential of the amputee (\ref{amputee}):
\begin{equation}
\label{t-exponential}
T[e^{A}](\phi)=e^{\frac{1}{2}\phi D^{-1}_0\phi}\sqrt{\det{\frac{D_0}{2\pi}}}\int\mathcal{D}_\Lambda J\,\,\,e^{W(J)+iJ\phi}.
\end{equation}
To obtain the right hand side, just use the definition (\ref{amputee}) and make the required substitution in the argument of the amputee inside the integral. 

Now recall eq. (\ref{Fourier}): $\exp{W(J)}$ is the Fourier transform of $\exp{\big(-S_\Lambda(\phi)\big)}$, and in the right hand side of eq. (\ref{t-exponential}) I have the inverse Fourier transform. I conclude that
\begin{equation}
\label{logarithm}
\log{T[e^A]}(\phi)=\frac{1}{2}\phi D^{-1}_0\phi - S_\Lambda(\phi)+const.
\end{equation}

Now apply $P$ to both sides of this equation, use eqs. (\ref{projector}), (\ref{constant}), and (\ref{propagator}) to obtain
\begin{equation}
\label{almost}
P\log T[e^A](\phi)=0.
\end{equation}

Now  apply $PT^{-1}$ to both sides of this equation and add to it the equation $PT^{-1}(1-P)\log T\exp A =0$, which holds because of the requirement (\ref{requirement}). You obtain
\begin{equation}
\label{inaction-eq}
PT^{-1}\log T\exp A = 0.
\end{equation}
This is called in \cite{Pivovarov:2009wa} the inaction equation.

Now recall that $A=I_P+V_P$ (eqs. (\ref{inaction}) and (\ref{vertex})) . I am going to argue that the inaction equation defines $V_P$ in terms of $I_P$. Indeed, this is a vectorial equation, and there is an  equation for each dimension in the range of the projector $P$, and $V_P$ belongs to the same range of $P$. So, the number of equations equals the number of unknowns. Now, compute the left hand side of the inaction equation in the linear approximation in $A$:
\begin{equation}
\label{linear}
PT^{-1}\log T\exp A \approx PT^{-1}\log (1+TA)\approx PT^{-1}TA=PA=V_P.
\end{equation}

So, I got an equation of the form
\begin{equation}
\label{new-equation}
V_P=\Phi[I_P+V_P],
\end{equation}
where the transform $\Phi$ is defined as
\begin{equation}
\label{Phi}
\Phi \equiv P(1-T^{-1}\log T\exp).
\end{equation}

Eq. (\ref{new-equation}) is almost what I wanted, the only difference with respect to eq. (\ref{my-transform}) is the presence of $V_P$ in its right hand side. This can be mended. Indeed, expansion of $\Phi[X]$ in $X$ at zero starts at a term of order $X^2$. So, approximately, up to terms cubic in $I_P$, I can neglect $V_P$ in the right hand side of eq. (\ref{new-equation}).

This can be used as a start of an iteration process. For example, on the second step of the iteration process I got
\begin{equation}
\label{iteration}
V_P=\Phi[I_P+\Phi[I_P]]+O(I_P^4).
\end{equation}

Generally, I have
\begin{equation}
\label{general}
V_P=\{(1-P+\Phi)^N - (1-P)(1-P+\Phi)^{N-1}\}[I_P] +O(I_P^{N+2}).
\end{equation}
Here the powers are understood as combinations of the transforms: the second power of a transform means that the transform is applied to the outcome of its first use. Notice that the second term in the curly brackets can be simplified as follows: $(1-P)(1-P+\Phi)^{N-1}[I_P]=I_P$. This is due to the presence of the projector in the right hand side of the definition (\ref{Phi}), and I recall that $PI_P=0$. Also the first term in the curly brackets is $(1-P+\Phi)^N=(1-PT^{-1}\log T\exp)^N$.

The above reasoning motivates me to \emph{define} the quantum transform as the following limit:
\begin{equation}
\label{definition}
\mathcal{Q}_P = \lim_{N\rightarrow\infty}\big[(1-PT^{-1}\log T\exp)^N-1\big].
\end{equation}
(I warn the reader against the temptation to use the binomial formula for expanding the power under the limit: this is a power of a nonlinear transform and the binomial formula does not work.)

The limit in the definition of the quantum transform exists at least in the sense of the perturbation theory: If at fixed $N$ I apply the mapping in the right hand side to an argument and expand it in powers of the argument, a coefficient at fixed power of the argument will in general be dependent on $N$. But it stops to change at $N$ large enough. For future reference, if I want to compute $V_P$ up to $n$th power in $I_P$, it suffices to take $N=n-1$ in eq. (\ref{definition}) for obtaining the right answer, and the expansion starts from $n=2$.

Comparison with conventional perturbative renormalization reveals that each iteration in eq. (\ref{definition}) generates a layer of subgraphs, and the projector at each layer makes the subtractions. Obviously, the renormalization scheme that agrees perturbatively with eq. (\ref{definition}) is defined by the conditions $(1-P)V_P=0$, which are automatically satisfied by the vertex functional. If nothing else, eq. (\ref{definition}) is an economical representation of the combinatoric intricacies of the perturbative renormalization. 

To summarise, here are the rules of this game: Take any unital algebra for which  the exponential and logarithm are defined. Pick up a projector with a finite dimensional kernel acting in the algebra. Make a choice of $T$---an invertible linear operator leaving the kernel of the projector invariant---and see if the limit in eq. (\ref{definition}) exists. If it does, study the dependence of the limit on $T$. If the limit stays in place even if you send $T$ to infinity in a certain way (modeling the UV divergences), you got a nontrivial quantum theory renormalized to the very end. That is, with all the parameters extracted from the correlators.

I leave this game for future studies. My next purpose is testing the above formalism on a simple but nontrivial example.

\section{Combinatorics and UV convergence}
\label{Combinatorics}

If I consider a theory with bare action which is a polynomial of the $m$-th power in the fields ($m>2$), all the conditions of Section \ref{Setup} will be satisfied for the projector nullifying completely all the powers of the fields involved in the bare action, and the ``free'' propagator coinciding with the exact propagator:
\begin{equation}
\label{projector-n}
P[A](\phi)=P_n[A](\phi)\equiv A_n(\phi),\,\,D_0=D.
\end{equation}
Here $A_n(\phi)$ is $A(\phi)$ with the starting part of its expansion in the fields subtracted up to power $n\leq m$ inclusively, and $D$ is the complete propagator, $W(J)=-JDJ/2+O(J^3)$.

In this case, $V_P$ is the generating functional of connected vertexes with more than $n$ external legs, which are amputated, and $I_P$ is the same but with the number of external legs not exceeding $n$. Now, the formalism of the previous section allows to express the amplitudes with more than $n$ particles involved as a series in the amplitudes with the number of particles not exceeding $n$. (Such expressions may be of interest for description of the processes with many particles involved.)

Considering this perturbation theory provided by powers of the transform $(1-PT^{-1}\log T\exp)$, I see that nothing but the conventional Feynman diagrams appear, with the lines corresponding to the exact propagator, and vertexes, to the exact amplitudes with number of external lines not exceeding $n$. 

In this case, however, the rules of computing the symmetry coefficients by each diagram differ the conventional ones, because each diagram appears several times. Each appearance of an individual diagram corresponds to a certain level of iteration, and can be related to a particular way to single out subgraphs in the diagram. The process of computing this generalized symmetry factors mimics the familiar procedure of UV renormalization, with the substantial difference that the subtraction procedure simply drops subgraphs with the number of external lines exceeding $n$.

This exercise for $n=4$ and $I_P$ even in the fields is performed in \cite{Kim&Pivovarov}. The expansion of $V_P$ up to the fourth order in $I_P$ is given there. It has required, as explained in the previous Section, considering the cube of the transform $(1-PT^{-1}\log T\exp)$. 

It is remarkable in this answer that not only the disconnected graphs have not appeared in it (this was expected because of the presence of the logarithm in the definition of the quantum transform), but also all the graphs UV divergent in four dimensions (there are altogether 15 of them) have completely canceled out.

I stress that the computation of the symmetry coefficients is a strictly combinatorial matter, and does not use any properties of the Feynman integrals. So, it appears that the combinatorics involved here knows the dimension of the space-time.

In my view, this result constitutes an important consistency check of the inaction approach.

Formulating the rules for computing the symmetry coefficients for the Feynman diagrams generated by the quantum transform is an interesting and important problem. It is addressed in \cite{Kim&Pivovarov}.

\section{Conclusions}
\label{Conclusions}

The inaction approach laid out above is an abstraction of the successful renormalization practice of perturbative quantum field theories. The abstraction allows one to give this practice a non-perturbative meaning embodied in the quantum transform. It also reveals that the raison d'etre of renormalization is not to overcome UV divergences, but to describe a complicated object---a logarithm of a Fourier transform of a function possessing extra properties---in terms of itself. Understood in this way, the inaction approach may be used in the areas superficially not related to quantum field theory, where such complicated objects appear. 
To give an example, the inaction approach can be used in probability theory for describing the properties of the so-called second characteristic function of a probability distribution (see \cite{Lukacs} for a definition of this object).

On a more pragmatic note, one could ask for a list of the benefits the new approach promises in exchange for the betrayal of the long and successful tradition of using the action functional as the key object in describing quantum systems. If the hopes do not count, I still can point out the undeniable fact that the inaction approach gives its user the freedom of choice between infinitely many possibilities to decide what to describe and in which terms. One example of putting this freedom to a good use was given in Section \ref{Combinatorics}.

A number of questions naturally arise within the new approach. The first one is, probably, on the interplay between symmetries of the theory and the inaction formulation. An attempt to address this issue see in \cite{Pivovarov:2012wz}. But the appearance of quantum anomalies had not been considered there.

The second question I want to point out is on the relationship between the renormalization group and the inaction approach. There is a subtle point here. One can obtain certain renormalization group equations withing the inaction approach considering a family of continuously parametrised projectors. This has been done in \cite{Pivovarov:2009wa}.  But there is also another possibility. One can consider continuous interpolation between projectors with the ranges of different dimensions. Notice, that if the projector in the definition of the quantum transform, eq. (\ref{definition}), is replaced with zero operator, the quantum transform vanishes, which corresponds to putting all the theory into the inaction functional. If it would be possible to continuously increase the range of the projector starting from zero,
a new kind of perturbation theory in the dimension of the space of unknowns would become possible.  

Next question is about the existence of the limit in eq. (\ref{definition}) beyond perturbation theory. It may be of interest to consider this limit for finite dimensional matrix algebras. Albeit the physical interpretation in this case is not clear (the correlators take values not in complex numbers but in matrices), such a consideration may shed some light on the possibility of existence of the quantum transform beyond perturbation theory. 

The biggest question of all is why the particular mapping of eq. (\ref{definition}) describes the Nature. What are its defining properties? It is probably better to put this big one aside, and invest efforts in solving the more concrete problems above.

\acknowledgments

I wrote this paper during a visit to CERN Theory division. I want to thank its staff for the hospitality, and the Ministry of Education and Science of the Russian Federation for supporting this visit. I also thank Dmitry Gorbunov, Victor Kim, and Sergey Trunov for helpful discussions.

\end{document}